\documentclass[a4paper]{article}

\usepackage{INTERSPEECH2020}

\title{DiDiSpeech: A Large Scale Mandarin Speech Corpus}
\name{Tingwei Guo, Cheng Wen, Dongwei Jiang, Ne Luo, Ruixiong Zhang, Shuaijiang Zhao, \\
	Wubo Li, Cheng Gong, Wei Zou, Kun Han, Xiangang Li}
\address{DiDi Chuxing, Beijing, China}
\email{\{guotingwei,wencheng,jiangdongwei,luone,zhangruixiong,zhaoshuaijiang,liwubo,gongcheng, zouwei,kunhan,lixiangang\}@didiglobal.com}

\begin{document}

\maketitle

\begin{abstract}
This paper introduces a new open-sourced Mandarin speech corpus, called DiDiSpeech. It consists of about 800 hours of speech data at 48kHz sampling rate from 6000 speakers and the corresponding texts. All speech data in the corpus is recorded in quiet environment and is suitable for various speech processing tasks, such as voice conversion, multi-speaker text-to-speech and automatic speech recognition. We conduct experiments with multiple speech tasks and evaluate the performance, showing that it is promising to use the corpus for both academic research and practical application. The corpus is available at https://outreach.didichuxing.com/research/opendata/. 
\end{abstract}

\section{Introduction}
\label{sec:intro}
Speech processing has been actively studied for decades. Powered by the emerging technology on deep learning, rapid progress has been made on the key technologies in speech processing \cite{DBLP:conf/icassp/ShenPWSJYCZWRSA18, DBLP:journals/corr/abs-1905-09263}, including automatic speech recognition (ASR) \cite{hinton2012deep,chorowski2015attention, DBLP:conf/icassp/ChanJLV16, DBLP:conf/iscslp/ZouJZYL18}, text-to-speech (TTS) \cite{DBLP:conf/icassp/ShenPWSJYCZWRSA18, DBLP:journals/corr/abs-1905-09263, DBLP:journals/corr/abs-2005-05106}, speaker identification (SID) \cite{DBLP:conf/icassp/SnyderGSPK18, DBLP:conf/icassp/LeiSFM14}, speech emotion recognition \cite{DBLP:conf/interspeech/HanYT14, DBLP:conf/icassp/MirsamadiBZ17, litmt}.

In recent years, due to the increasing research in the speech community, many speech datasets are published for the usage of research and industry. However, limited by the quality or quantity of the speech corpora, improvements in research community may not always scale up to industrial scenarios. Taking Mandarin speech synthesis as an example, the difficulty of building high-quality single-speaker TTS models has been greatly reduced with the proposed effective neural TTS algorithms \cite{DBLP:journals/corr/abs-1905-09263, DBLP:journals/corr/abs-2005-05106}. The focus of speech synthesis is changing from synthesizing human-like speech of a given target speaker to more complex tasks, such as voice conversion (VC) \cite{ DBLP:conf/interspeech/KanekoKTH19} and multi-speaker TTS \cite{DBLP:conf/nips/JiaZWWSRCNPLW18}. A number of corpora (e.g., VCTK \cite{yamagishi2019cstr}, VCC 2018 \cite{lorenzo2018voice} and LibriTTS \cite{zen2019libritts}) are available for this area, whereas all of them only contain recordings in English. A few Mandarin speech corpora such as THCHS30 \cite{Wang2015THCHS}, HKUST \cite{DBLP:conf/iscslp/LiuFYCHG06}, AISHELL-1 \cite{bu2017aishell} and CSMSC \cite{data2017csmsc} have also been released. However, these datasets are designed for either ASR or single-speaker TTS and are not suitable for other speech tasks \cite{DBLP:journals/corr/abs-2007-14602}, such as multi-speaker TTS or VC.

Considering these limitations, we release a novel Mandarin speech database, called DiDiSpeech, which is designed for various speech processing tasks including ASR, TTS, SID, etc. DiDiSpeech consists of two parts: DiDiSpeech-1 and DiDiSpeech-2. The DiDiSpeech-1 is a 572-hour Mandarin speech corpus, which is composed of both the parallel corpus (sentences uttered by all speakers with the same content) and non-parallel corpus of 4500 speakers. The DiDiSpeech-2 consists of sentences from 1500 speakers and each of the speakers recorded more than 100 non-parallel utterances.

The rest of this paper is organized as follows: Section 2 summarizes the publicly available speech corpora; Section 3 introduces the details of the DiDiSpeech and the data processing pipeline used to produce this corpus; Section 4 shows the experimental results conducted on the DiDiSpeech, and Section 5 concludes this paper.

\begin{table}[htp]
	\caption{List of publicly available speech corpora.}
	\label{tab:publicly_speech_corpora}
	\setlength{\tabcolsep}{5pt} %% default is 6pt
	\centering
	\begin{tabular}{llccc}
		\toprule
		\textbf{Corpus}	 & \textbf{Language} & \textbf{Hours} & \textbf{Total}	& \textbf{Sampling} \\
		& &&\textbf{Speakers}&\textbf{rate(kHz)} \\
		\midrule
		VCTK & English & 44 & 109 & 48 \\ % inserting body of the table
		VCC 2018 & English & 1 & 12 & 22.05\\
		LibriTTS & English & 585 & 2456 & 24\\
		CSMSC & Mandarin & 12 & 1 &48\\
		HKUST & Mandarin & 200&2100 & 8\\
		AISHELL-1 & Mandarin & 170 & 400& 16\\
		DiDiSpeech-1 &  Mandarin & 572 & 4500 & 48\\
		DiDiSpeech-2 &  Mandarin & 227 & 1500 & 48\\
		\bottomrule
	\end{tabular}
\end{table}

\section{Related Work}
\label{sec:Related Work}

In recent years, numerous open-source speech databases have been released. Table \ref{tab:publicly_speech_corpora} lists the details of the publicly available speech corpora.

For English speech synthesis, VCTK \cite{yamagishi2019cstr} and VCC 2018 \cite{lorenzo2018voice} have been widely used in the research of multi-speaker TTS and VC respectively. The VCTK corpus consists of 44 hours of speech data uttered by 109 native speakers of English with various accents. In VCC 2018, the voices of only 12 speakers are recorded, while each speaker reads out around 80 sentences. To ensure the quality of speech data, all voices in two corpora are recorded in semi-anechoic chambers. However, limited by the recording cost, the scale of both them cannot meet the needs of researchers. In contrast, the LibriTTS \cite{zen2019libritts} database provides a large amount of multi-speaker speech data that can be used in speech synthesis tasks. It contains more than 585 hours English speech data recorded from 2456 speakers. However, the LibriTTS is a subset of the LibriSpeech \cite{DBLP:conf/icassp/PanayotovCPK15} database which is designed for English ASR tasks. For Mandarin speech synthesis, the only freely available corpus is CSMSC, containing 12 hours of speech data of only one speaker.

For ASR, several open-source read speech corpora, such as HKUST and AISHELL-1, have also been released. Since the scale of these databases is significantly smaller than the ASR datasets used in companies \cite{DBLP:journals/corr/abs-2005-09862, DBLP:journals/corr/abs-1910-09932}, more data is still needed to bridge the technology gap between research and industry.

\section{DiDiSpeech Database}
\label{sec:DiDiSpeech Database}
This section describes overall statistics of the DiDiSpeech, and the data processing pipeline used to produce the database.
\subsection{Speaker information}
\label{sec:Teacher}
\begin{figure}[h]
	\centering
	\includegraphics[width=1.0\linewidth]{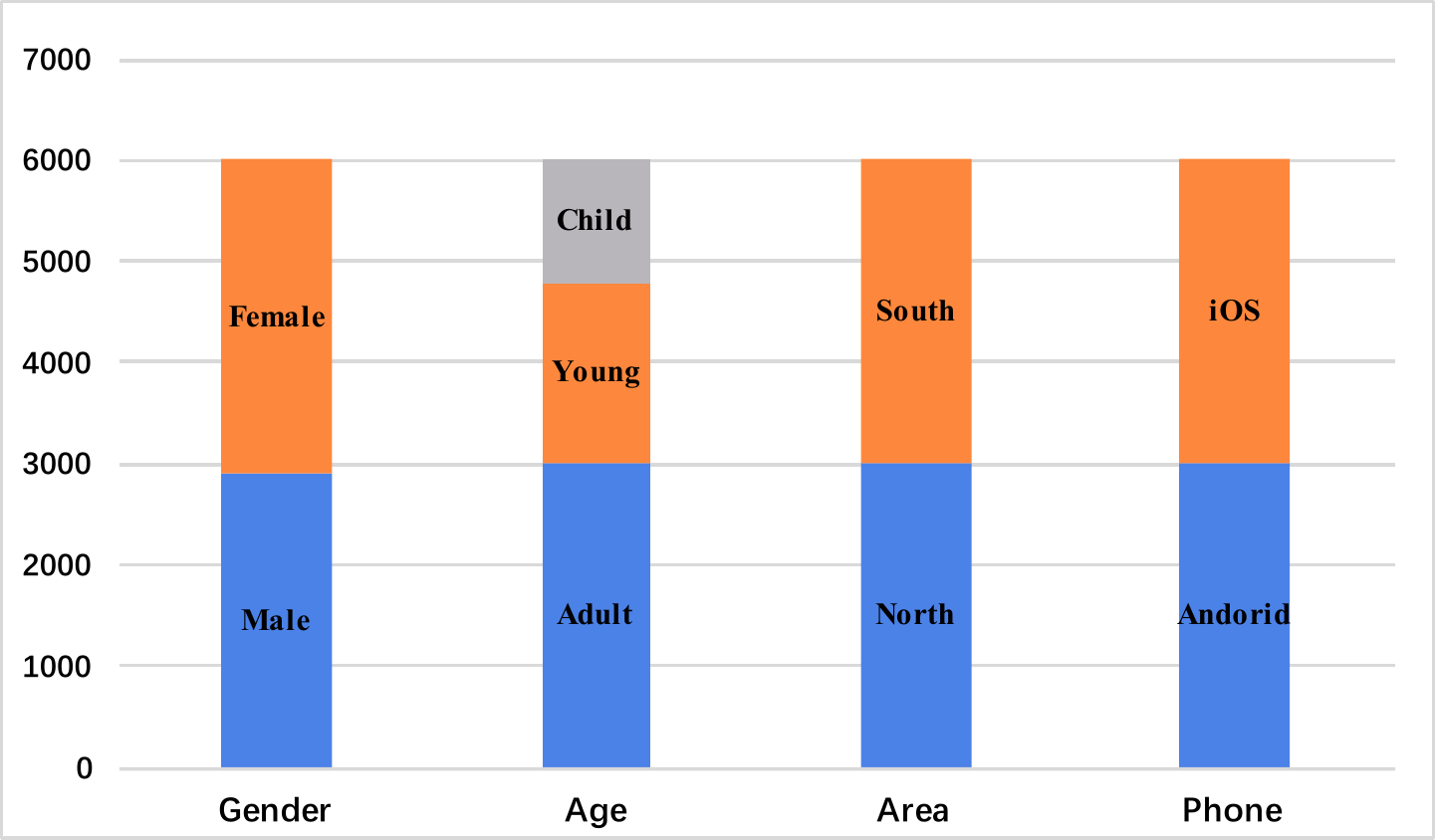}
	\caption{Speaker statistics.}
	\label{fig:speaker_information}
\end{figure}

Figure \ref{fig:speaker_information} illustrates the detailed information of speakers in DiDiSpeech database. There are 6000 subjects aged from 6 to 79 in the corpus. All subjects are native Mandarin speakers with no or slight accent. Taking the Qinling and Huaihe River as the dividing line, the number of speakers from north to south is roughly equal.

In order to meet the needs of various tasks, 6000 speakers are divided into two parts to perform voice recording, which results in two different subsets of DiDiSpeech, called DiDiSpeech-1 and DiDiSpeech-2 respectively. Table \ref{tab:speaker_gender_information} and Table \ref{tab:speaker_area_information} show the gender and geographical distribution statistics of speakers in both of them.

\begin{table}[h]
	\caption{Speaker Gender Information.}
	\label{tab:speaker_gender_information}
	\setlength{\tabcolsep}{10pt} %% default is 6pt
	\centering
	\begin{tabular}{ccccc}
		\toprule
		& \multicolumn{2}{c}{\textbf{DiDiSpeech-1}}& \multicolumn{2}{c}{\textbf{DiDiSpeech-2}}  \\
		%%& \textbf{DiDiSpeech-1} && \textbf{DiDiSpeech-2}	&  \\
		\cline{2-5}
		\specialrule{0em}{2pt}{1pt}
		\textbf{Gender}	 & \textbf{Male} & \textbf{Female} & \textbf{Male}	& \textbf{Female} \\
		\midrule
		Child & 383 & 505 & 195 & 120 \\ % inserting body of the table
		Youth & 652 & 668 & 258 & 207 \\
		Adult & 1059 & 1233 & 344 & 376 \\
		\bottomrule
	\end{tabular}
\end{table}

DiDiSpeech-1 is aimed for VC task. Both the parallel corpus and the non-parallel corpus of 4500 speakers are supported in DiDiSpeech-1. Each speaker has 50 parallel sentences, while the remaining more than 50 sentences are different. Among all speakers in DiDiSpeech-1, the ratio of male to female is close to 1:1. In terms of the distribution of speakers in age, adults, youths and children account for 50\%, 30\% and 20\% of the total respectively, while adults refer to people aged 20 and over, youths refer to people aged 13 to 19, and children are people under 13 years of age.

\begin{table}[h]
	\caption{Speaker Area Information.}
	\label{tab:speaker_area_information}
	\setlength{\tabcolsep}{10pt} %% default is 6pt
	\centering
	\begin{tabular}{ccccc}
		\toprule
		& \multicolumn{2}{c}{\textbf{DiDiSpeech-1}}& \multicolumn{2}{c}{\textbf{DiDiSpeech-2}}  \\
		\cline{2-5}
		\specialrule{0em}{2pt}{1pt}
		\textbf{Area}	 & \textbf{North} & \textbf{South} & \textbf{North}	& \textbf{South} \\
		\midrule
		Child & 474 & 414 & 214 & 101 \\ % inserting body of the table
		Youth & 653 & 667 & 244 & 221 \\
		Adult & 1106 & 1186 & 288 & 432 \\
		\bottomrule
	\end{tabular}
\end{table}

In DiDiSpeech-2, 1500 subjects, of which 797 are male and 703 are female speakers, are arranged to record more than 100 non-parallel sentences. The DiDiSpeech-2 is established for the purpose of building multi-speaker speech synthesis or ASR system. In order to increase the contextual and phonetic coverage, the corpus of each speaker is selected based on a greedy algorithm from a large basic corpus. 

All speech data in DiDiSpeech is recorded by using mobile phones in quiet environment. The recording is made on device with the phone operating in hands-free mode, while all speakers are prompted by printed scripts. As shown in Figure 1, the ratio of speakers using Android and iOS phones is about 1:1. All audio in DiDiSpeech is saved in an uncompressed waveform, with a sampling rate of 48KHz and a bit depth of 16bit.

\subsection{Text pre-processing}
\label{sec: Text Processing}
The sentences in DiDiSpeech are mainly selected from newspapers and books. The number of characters in a single sentence does not exceed 36. In DiDiSpeech-1, 225000 parallel utterances and 255571 non-parallel utterances have been uttered by 4500 speakers, while each sentence contains an average of 14 characters. DiDiSpeech-2 contains 171361 non-parallel utterances from 1500 speakers and 2701693 characters in total. In order to reduce the recording errors caused by text ambiguity, non-standard words (such as numerals, dates, etc.) in sentences have been detected and normalized.

In addition, a grapheme-to-phoneme (G2P) conversion module is adapted to generate pronunciations for Mandarin sentences, where a transformation-based learning algorithm is employed for polyphone disambiguation \cite{DBLP:conf/interspeech/PolyakovaB06}. After that, we also list the polyphonic words with correct pronunciations searched from a manual-tagging corpus into a dictionary to further repair the results of G2P.

\begin{figure}[ht]
	\centering
	\includegraphics[width=1.0\linewidth]{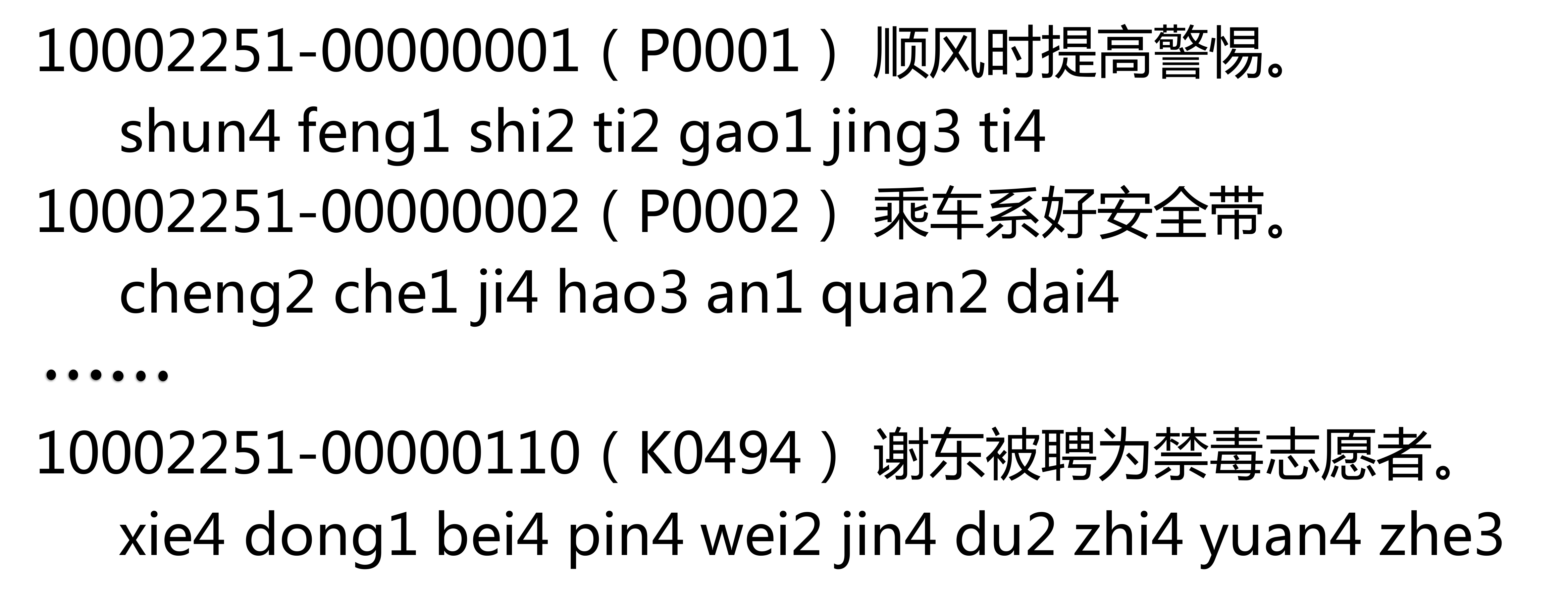}
	\caption{Text annotation example.}
	\label{fig:text_annotation_example}
\end{figure}

Figure \ref{fig:text_annotation_example} shows the format in which the text and pronunciations are organized. The sentences of each speaker are recorded in a single file. The parallel sentences and non-parallel sentences in DiDiSpeech-1 are marked with P and K respectively.

\subsection{Post processing}
\label{sec:  Audio Processing}
We have carried out a statistical analysis of signal-to-noise ratio (SNR) of speech data in DiDiSpeech to filter out the audio with significant background noise. In addition, on the basis of energy and front-end prosody analysis, the waveform with abnormal pauses is detected and screened out.

Furthermore, the DiDi speech recognition engine is applied to search utterances with audio/text mismatches in DiDiSpeech, which can be caused by reader-introduced insertions, deletions, substitutions, and transpositions. These sentences are manually inspected before being deleted from DiDiSpeech.

\section{Experiments}
\label{sec:experiments}
This section presents experimental results for several speech processing tasks on the DiDiSpeech dataset.
\subsection{Voice conversion}
% model structure
For this paper, we use StarGan-VC2 framework proposed in \cite{DBLP:conf/interspeech/KanekoKTH19} for both Parallel and Non-Parallel VC experiments. We design the network architectures on the basis of CycleGAN-VC2, i.e., we use a 2-1-2D convolutional neural networks (CNN) in generator and a 2D CNN in discriminator. We formulate discriminator using the projection discriminator. In the pre-experiment, we find that skip connections in residual blocks result in partial conversion. Thus, we remove them in generator. For a generative adversarial networks (GAN) objective, we use a least squares GAN. We conduct speaker-wise normalization for a pre-process. We train the networks using the Adam optimizer with a batch size of 8, in which we use a randomly cropped segment (128 frames) as one instance. The number of iterations is set to 3 $\times$ 105, learning rates for generator and discriminator are set to 0.0002 and 0.0001, respectively, and the momentum term is set to 0.5. We set $\lambda_{cyc}$ = 10,
$\lambda_{id}$ = 5, and $\lambda_{cls}$ = 1. We use $L_{id}$ only for the first 104 iterations to stabilize the training at the beginning.

% data processing
For voice conversion experiments, we take speech data of 500 speakers randomly chosen from DiDiSpeech-1. Note that only the first 50 sentences of each speaker in DiDiSpeech-1 is parallel, so the number of pairs for parallel training is 50 $\times$ 500 = 25000. For non-parallel voice conversion, we randomly choose from non-parallel parts of these 500 speakers in DiDiSpeech-1, resulting in 50 $\times$ 500 = 25000 pairs. The recordings of DiDiSpeech-1 are downsampled to 16kHz for this experiment. We extract 36-dimensional mel-cepstral coefficients (MCEPs), logarithmic fundamental frequency (log F0), and aperiodicities (APs) every 5ms by using the WORLD \cite{DBLP:journals/ieicet/MoriseYO16} analyzer. After training, we also use WORLD as the vocoder. 

We use Mean Opinion Score (MOS) on a scale from 1.0 to 5.0 as the measuring metric of our parallel and non-parallel VC models. The results of the MOS tests are shown in Table \ref{tab:vc_results}. Samples of synthesized audio are available at GitHub\footnote[1]{https://athena-team.github.io/DiDiSpeech/}.

\begin{table}[ht]
	\caption{Speech naturalness and speaker similarity Mean Opinion Score evaluations with 95\% confidence
		intervals for VC task (higher is better).}
	\label{tab:vc_results}
	\setlength{\tabcolsep}{12pt} %% default is 6pt
	\centering
	\begin{tabular}{lcc}
		\toprule
		& \textbf{Naturalness} & \textbf{Similarity} \\
		\midrule
		Ground Truth & 4.17${\pm}$0.09 & - \\
		Parallel VC &  3.37${\pm}$0.07 & 3.43${\pm}$0.06 \\
		Non-Parallel VC &  3.23${\pm}$0.07 & 3.27${\pm}$0.05 \\
		\bottomrule
	\end{tabular}
\end{table}

\subsection{Multi-speaker speech synthesis}
We build two multi-speaker speech synthesis systems for DiDiSpeech-2 dataset. The first one is similar to \cite{DBLP:conf/nips/JiaZWWSRCNPLW18}. The system is composed of three components, including a speaker encoder, a sequence-to-sequence synthesizer and a vocoder. The second one is a FastSpeech \cite{DBLP:journals/corr/abs-1905-09263} based multi-speaker speech synthesis system.

Next, we describe the detailed architecture of our first multi-speaker speech synthesis system. We conduct the experiments on audio that downsampled to 24kHz. The speaker encoder is a pre-trained model\footnote[2]{http://kaldi-asr.org/models/m8 (SITW Xvector System 1a)} for the speaker verification task using kaldi trained on VoxCeleb1\&2 \cite{DBLP:journals/csl/NagraniCXZ20, DBLP:conf/interspeech/ChungNZ18}, and is able to extract a 512-dimensional x-vector embedding from a reference audio. In the embedding space, the embeddings computed from audio of same speaker are close and which from different speakers are supposed to be far. The synthesizer architecture is Tacotron-2 \cite{DBLP:conf/icassp/ShenPWSJYCZWRSA18}, and the hyper-parameters setting is similar to the original paper. The difference is that we use a reduction factor of 2 and a sigmoid layer with weighted binary cross entropy loss for the prediction of "stop token", in which the weight of positive examples is set to be 20. Besides, both the L1 and Mean Square Error (MSE) losses are used for the acoustics features prediction. The inputs of the synthesizer are sequences of phonemes mapped from pronunciations and target outputs are normalized 80-dimensional log mel-spectrograms extracted from audio with a hop length of 256 and a window size of 1024. During the training and inference phases of our system, the extracted speaker embeddings are concatenated with the synthesizer's encoder output and passed to the attention layer in order to achieve multi-speaker speech synthesis, while the parameters of the speaker encoder are frozen. 

We use multi-band MelGAN \cite{DBLP:journals/corr/abs-2005-05106} as vocoder to convert mel-spectrograms to audio. The analysis filters and synthesis filters proposed in \cite{yu2019durian} are employed for multi-band processing. A single generator is adopted to predict 4 sub-band signals simultaneously, where upsampling layers with 8x, 4x and 2x factors respectively are used to achieve 256x upsampling.

\begin{table}[t]
	\caption{Speech naturalness Mean Opinion Score evaluations with 95\% confidence
		intervals for multi-speaker speech synthesis (higher is better).}
	\label{tab:multispk_tts_mos_results}
	\setlength{\tabcolsep}{12pt} %% default is 6pt
	\centering
	\begin{tabular}{lcc}
		\toprule
		& \textbf{Seen Speakers} & \textbf{Unseen Speakers} \\
		\midrule
		Ground Truth &	4.11${\pm}$0.08 &	4.20${\pm}$0.10 \\
		Tacotron-2 &  3.72${\pm}$0.07 & 3.60${\pm}$0.08 \\
		FastSpeech &  3.58${\pm}$0.07 & 3.53${\pm}$0.09 \\
		\bottomrule
	\end{tabular}
\end{table}

\begin{table}[t]
	\caption{Speaker similarity Mean Opinion Score evaluations with 95\% confidence
		intervals for multi-speaker speech synthesis (higher is better).}
	\label{tab:multispk_tts_simliarity_results}
	\setlength{\tabcolsep}{12pt} %% default is 6pt
	\centering
	\begin{tabular}{lcc}
		\toprule
		& \textbf{Seen Speakers} & \textbf{Unseen Speakers} \\
		\midrule
		Tacotron-2 &  3.78${\pm}$0.06 & 3.61${\pm}$0.06 \\
		FastSpeech &  3.54${\pm}$0.07 & 3.34${\pm}$0.10 \\
		\bottomrule
	\end{tabular}
\end{table}

In our second multi-speaker speech synthesis system, a FastSpeech model is employed to replace the Tacotron-2 synthesizer. For the FastSpeech model, the dimension of phoneme embeddings, the hidden size of the self-attention and 1D convolution in the FFT block are all set to 384. The number of attention heads is set to 2. The kernel sizes of the 1D convolution in the 2-layer convolutional network are both set to 3, with input/output size of 384/1536 for the first layer and 1536/384 for the second layer. The dimension of speaker embedding is 512 and it is applied both before sending into the encoder and after we get output of encoder. The output linear layer converts the 384-dimensional hidden layer output into 80-dimensional mel-spectrogram. In duration predictor, the kernel sizes of the 1D convolution are set to 3, with input/output sizes of 384/384 for both layers. The FastSpeech speech synthesis system is built using Athena\footnote[3]{https://github.com/athena-team/athena}.

We run a MOS hearing test to measure the performance of our multi-speaker speech synthesis systems on both seen speakers and unseen speakers. Speech naturalness and speaker similarity MOS are reported in Table \ref{tab:multispk_tts_mos_results} and Table \ref{tab:multispk_tts_simliarity_results}. Samples of synthesized audio are available at GitHub\footnote[4]{https://athena-team.github.io/DiDiSpeech/}.

\subsection{Automatic speech recognition}
\label{ssec:asr}
We conduct the experiments of ASR task on our DiDiSpeech-2 dataset. As shown in Table \ref{tab:ASR_dataset}, we randomly split the DiDiSpeech-2 dataset as the training set (85\%), development set (15\%) and test set (5\%). In the audio preprocessing, all speech data are downsampled to 16kHz. The input of ASR models is filter-bank features extracted from 25ms frames with 10ms window shift. We use characters as labels to train and evaluate all ASR models.

\begin{table}[ht]	
	\caption{Summary of DiDiSpeech-2 dataset for ASR task.}
	\label{tab:ASR_dataset}
	\setlength{\tabcolsep}{10pt} %% default is 6pt
	\centering
	\begin{tabular}{lccc}
		\toprule
		& \textbf{Training set} & \textbf{Dev set } & \textbf{Test set}  \\
		\midrule
		Utterances     & 145661          & 17990               & 7710          \\ 
		Speakers	& 1500	&1500	&1494 \\ 
		\bottomrule
	\end{tabular}
\end{table}

Recently, Transformer-based ASR models have achieved competitive performance on HKUST, AISHELL-1 and LibriSpeech \cite{DBLP:journals/corr/abs-2005-09862,DBLP:journals/corr/abs-1910-09932,DBLP:conf/icassp/LiuYCHL20}. In the ASR experiments, we follow model structure of previous work \cite{DBLP:journals/corr/abs-1910-09932} with ${e}$ = 12, ${d}$ = 6, ${d_{model}}$ = 512, ${d_{ff}}$ = 1280 and ${d_{head}}$ = 8. We adopt downsampling before Transformer encoder, resulting in a 4-fold downsample in total. For training steps, all models are trained on 4 GPU with a total batch size of 128 for 300k steps. We use the Adam optimizer \cite{DBLP:journals/corr/KingmaB14} and vary learning rate with warmup schedule \cite{DBLP:journals/corr/VaswaniSPUJGKP17}. For evaluating, the model with the lowest loss on development set is chosen. We also use beam search with beam size 10 and two layers long short-term memory (LSTM) language model while decoding. CTC-joint decoding is performed as proposed in \cite{DBLP:conf/icassp/KimHW17} with CTC weight 0.3. Table \ref{tab:ASR_result} shows the results decoding in test set of DiDiSpeech-2.
	
\begin{table}[ht]
	\caption{ASR results on test set of DiDiSpeech-2 (lower is better).}
	\label{tab:ASR_result}
	\setlength{\tabcolsep}{8pt} %% default is 6pt
	\centering
	\begin{tabular}{p{4cm}lc}
		\hline
		\toprule
		\textbf{Model}	&\textbf{CER(${\%)}$} \\
		\midrule
		Transformer-ASR & 4.05 \\
		+ Beam Search & 3.77 \\
		+ Language Model & 2.72 \\
		+ CTC & 2.50 \\
		\bottomrule
	\end{tabular}
\end{table}

\subsection{Paralinguistic information tasks}
This subsection describes the experimental setup and results for paralinguistic information tasks, 
including speaker classification, gender classification and age estimation. As shown in Table \ref{tab:datasets_splitation}, we divide the DiDiSpeech corpus into three parts (training set, development set and test set) by randomly selecting four utterances from each speaker for development set and test set respectively. The training and the testing are performed on same speakers. 

All audio is first downsampled to 16kHz and converted to filter-bank features with a frame length of 25ms and a frame shift of 10ms. 
Then, features are mean-normalized over a sliding window of 300 frames. 
No voice activity detection, silence removal processing, or data augmentation method is used. 

\begin{table}[t]
	\caption{Dataset statistics for speaker classification, gender classification and age estimation.}
	\label{tab:datasets_splitation}
	\setlength{\tabcolsep}{6pt} %% default is 6pt
	\centering
	\begin{tabular}{lcccc}
		\toprule
		& \textbf{Training set} & \textbf{Dev set} & \textbf{Test set} & \textbf{Total} \\
		\midrule
		Utterances & 603932 & 24000 & 24000 & 651932 \\
		Speakers & 6000 & 6000 & 6000 & 6000 \\
		\bottomrule
	\end{tabular}
\end{table}

\textbf{Speaker classification.} 
We fix the model architecture to be thin ResNet-34, which is similar to the standard ResNet-34 \cite{DBLP:conf/cvpr/HeZRS16}, except that it only contains 1/4 amount of channels to reduce computation.
We adopt 40-dimensional filter-bank features for the model.
Average pooling is used to obtain utterance-level embeddings. 
Then, a dense layer with 128 output units is followed by the pooling layer in order to resize the embeddings to the desired dimension.
The embeddings are then fed into a dense layer with softmax contains 6000 classes.
We adopt stochastic gradient descent optimizer with momentum 0.9, using a piecewise constant learning rate decay schedule. 
The learning rate values are [0.1, 0.01, 0.001], and decay boundaries are [20000, 50000] steps. 
The model is trained with a total batch size of 512 and trained for 50 epochs. 
During training, data in a mini-batch are randomly cropped to a fix length between [200, 1000] frames. 
%The model with the lowest loss on development set is used for evaluation. 
Results are presented in Table \ref{tab:paralinguistic_results}.

\textbf{Gender classification.} 
This is a three classes (adult males, adult females, and children) classification task. 
We adopt the same model architecture and optimizer as described in speaker classification.
To improve the accuracy, we adopt 80-dimensional filter-bank features.
The model is also trained on a total batch size of 128 and trained for 50 epochs.
Results are reported in Table \ref{tab:paralinguistic_results}.

\begin{table} [ht]
	\caption{Results for speaker classification and gender classification.
		Accuracies in percentage are reported (higher is better).}
	\label{tab:paralinguistic_results}
	\setlength{\tabcolsep}{6pt} %% default is 6pt
	\centering
	\begin{tabular}{lcccc}
		\toprule
		& \multicolumn{2}{c}{\textbf{Dev set}} & \multicolumn{2}{c}{\textbf{Test set}}  \\
		\cline{2-5}
		\specialrule{0em}{2pt}{1pt}
		& \textbf{Top-1} & \textbf{Top-5} & \textbf{Top-1} & \textbf{Top-5} \\
		\midrule
		Speaker Classification & 98.22 & 99.79 & 98.20 & 99.76  \\
		Gender Classification & 97.39 & - & 97.39 & - \\
		\bottomrule
	\end{tabular}
\end{table}

\begin{table} [ht]
	\caption{Results for age estimation.
		MAEs (lower is better) and Pearson's correlation coefficients $\rho$ ($\rho\in[0,1]$, higher is better) are reported.}
	\label{tab:age_results}
	\setlength{\tabcolsep}{8pt} %% default is 6pt
	\centering
	\begin{tabular}{lcccc}
		\toprule
		& \multicolumn{2}{c}{\textbf{Dev set}} & \multicolumn{2}{c}{\textbf{Test set}}  \\
		\cline{2-5}
		\specialrule{0em}{2pt}{1pt}
		& \textbf{MAE} & \textbf{$\rho$} & \textbf{MAE} &$\textbf{$\rho$}$ \\
		\midrule
		Age Estimation & 3.77 & 0.78 & 3.76 & 0.79 \\
		\bottomrule
	\end{tabular}
\end{table}

\textbf{Age estimation.}  
Age estimation consists of automatically determining the age of a speaker in a given segment of the speech utterance \cite{DBLP:conf/icassp/MinematsuSH02}.
We adopt 80-dimensional filter-bank features.
We also adopt the same model architecture and optimizer as described in speaker classification, except that Mean Absolute Error (MAE) and Pearson's correlation coefficient $\rho$ \cite{DBLP:journals/eaai/BahariMhL14} are used as objective measure to access the performance of the age estimation model.
It is worth noticing that we adopt a smaller initial learning rate of 0.01 to avoid non-convergence.
Results are reported in Table \ref{tab:age_results}.

\section{Conclusion}
\label{sec:Conclusion}
In this paper, we introduce DiDiSpeech, a large scale Mandarin speech corpus. The corpus includes about 800 hours speech data from 6000 speakers. Both audio and the corresponding texts are provided in the corpus. The experimental results show that the corpus is able to achieve good performances on multiple speech tasks, including VC, multi-speaker TTS, ASR and various paralinguistic information tasks. We expect the corpus to accelerate research and industry application in the speech community.

\bibliographystyle{IEEEtran}

\bibliography{refs}

% \begin{thebibliography}{9}
% \bibitem[1]{Davis80-COP}
%   S.\ B.\ Davis and P.\ Mermelstein,
%   ``Comparison of parametric representation for monosyllabic word recognition in continuously spoken sentences,''
%   \textit{IEEE Transactions on Acoustics, Speech and Signal Processing}, vol.~28, no.~4, pp.~357--366, 1980.
% \bibitem[2]{Rabiner89-ATO}
%   L.\ R.\ Rabiner,
%   ``A tutorial on hidden Markov models and selected applications in speech recognition,''
%   \textit{Proceedings of the IEEE}, vol.~77, no.~2, pp.~257-286, 1989.
% \bibitem[3]{Hastie09-TEO}
%   T.\ Hastie, R.\ Tibshirani, and J.\ Friedman,
%   \textit{The Elements of Statistical Learning -- Data Mining, Inference, and Prediction}.
%   New York: Springer, 2009.
% \bibitem[4]{YourName17-XXX}
%   F.\ Lastname1, F.\ Lastname2, and F.\ Lastname3,
%   ``Title of your INTERSPEECH 2020 publication,''
%   in \textit{Interspeech 2020 -- 20\textsuperscript{th} Annual Conference of the International Speech Communication Association, September 15-19, Graz, Austria, Proceedings, Proceedings}, 2020, pp.~100--104.
% \end{thebibliography}

\end{document}